\documentstyle[prd,aps]{revtex}
\begin{document}
\preprint{SNUTP 96-033;~ gr-qc/9608015}
\title{Renormalized Thermodynamic
Entropy of Black Holes in Higher Dimensions}

\author{Sang Pyo Kim$^a$ \footnote{Electronic
mail: sangkim@knusun1.kunsan.ac.kr},
Sung Ku Kim$^b$ \footnote{Electronic mail:
skkim@theory.ewha.ac.kr},
Kwang-Sup Soh$^c$ \footnote{Electronic mail:
kssoh@phyb.snu.ac.kr},
and Jae Hyung Yee$^d$ \footnote{Electronic mail:
jhyee@phya.yonsei.ac.kr}}

\address{$^a$ Department of Physics,
Kunsan National University,
Kunsan 573-701, Korea \\
$^b$ Department of Physics,
Ewha Womans University,
Seoul 120-750, Korea\\
$^c$ Department of Physics Education,
Seoul National University,
Seoul 151-742, Korea\\
$^d$ Department of Physics
and Institute for Mathematical
Sciences,
Yonsei University, \\
Seoul 120-749, Korea}

\maketitle
\begin{abstract}
We study the ultraviolet divergent structures of the matter
(scalar) field in a higher $D$-dimensional Reissner-Nordstr\"{o}m
black hole and compute the matter field contribution to the
Bekenstein-Hawking entropy by using the Pauli-Villars
regularization method. We find that the matter field
contribution to the black hole entropy does not, in general,
yield the correct renormalization of the gravitational coupling
constants. In particular we show that the matter field contribution
in odd dimensions does not give the term proportional to
the area of the black hole event horizon.
\end{abstract}


\section{Introduction}

One of the salient problems concerning the
Bekenstein-Hawking~\cite{bekenstein,hawking}
black hole entropy is to understand its
microscopic origin.
As an attempt to understand this problem
't~Hooft suggested that the thermodynamic entropy
of a scalar field coupled to a black hole
background can give rise to
the correct black hole entropy, provided
that a suitable brick wall is introduced
outside the event horizon~\cite{thooft}.
The entropy of the matter field
diverges as the brick wall approaches to
the event horizon, and Susskind and
Uglum~\cite{susskind} suggested that
the matter field contributions be interpreted
as the one-loop corrections to the classical
Bekenstein-Hawking entropy and renormalize
the gravitational coupling constants.
Demers, Lafrance and Meyers~\cite{demers} (DLM)
have subsequently confirmed,
by using the Pauli-Villars covariant regularization method,
the Susskind-Uglum conjecture for the scalar field
minimally coupled to a four-dimensional
Reissner-Nordstr\"{o}m (RN) black hole background.

DLM have shown that in four dimensions
the renormalization
constants arising in the renormalization of black hole
entropy are precisely the same as those arising
in the renormalization of the gravitational
action by a quantum (zero temperature)
scalar field. They used the Pauli-Villars
regularization by assuming that all
the regulator fields obey the Bose-Einstein
statistics. 

To further clarify the question on the matter field
contributions we
study, in this paper, the ultraviolet divergent structures of
a massive scalar field in a higher
$D$-dimensional RN black hole,
compute the matter field contribution to the
black hole entropy using the Pauli-Villars
regularization method, and
compare the results with
the divergent terms arising in the one-loop
renormalization of the gravitational action.

The thermodynamic entropy of a massive scalar field
in the $D$-dimensional RN black holes was already
found in the brick wall model~\cite{mann},
in which the RN black hole is linearized
as the Rindler form by considering a pill-box
shaped region close to the outer event horizon
and the number of states of the scalar field
in the Rindler metric is used to compute the entropy.
The $\zeta$-function regularization
scheme was also used to find the
divergent structure and
renormalized entropy of the scalar field
in a $D$-dimensional Rindler-like
spacetime~\cite{bytsenko}.
In this paper, however,
we consider the scalar field in the $D$-dimensional
RN black hole without approximating it by the Rindler form
near the outer event horizon,
find alternatively the number of states
from the semiclassical quantization of
Klein-Gordon equation, and get the divergent
structure of the free energy and entropy
as the divergent functions of brick wall
thickness.

In Sect. II we elaborate, by using the Pauli-Villars
regularization method, the one-loop renormalization
of the $D$-dimensional gravitational action
by a quantum scalar field. In the next three sections
we study the ultraviolet divergent structures of
the scalar field
and compute the scalar field contribution to the
$D$-dimensional RN black hole entropy by using
the Pauli-Villars regularization method
with all the regulator fields
treated as obeying the Bose-Einstein statistics.
We find in particular that the entropy contribution
of the scalar field minimally coupled to an
odd dimensional RN black hole background does not
give the term proportional to the surface area of the event
horizon. We conclude with some discussions in the last section.
The two appendices contain some mathematical formulas
needed for the computations in the main text.

Throughout this paper we adopt the
units, $c = \hbar = k = 1$, but keep the
gravitational constant $G$.
The spacetime signature is $(-, +, \cdots, +)$.

\section{Renormalization of Gravitational Action by
Quantum Fields}

Quantum field theory
has been extensively studied in curved spacetimes
(for references, see~\cite{birrel}).
Dimensional regularization, zeta-function
regularization, point-splitting, and Pauli-Villars
regularization methods have been developed
to find the renormalized effective action
for a quantum field.
We shall use the Pauli-Villars method that enables
us to regularize the thermodynamic entropy
of the quantum field in a black hole background
and to relate this directly with the
renormalization of action.
We shall further elaborate the Pauli-Villars regularization
method efficient
in evaluating the effective action and
the thermodynamic entropy.

The one-loop effective action
for gravity in $D$ dimensions~\cite{birrel} takes the form
\begin{equation}
{\cal I}_D = \int d^Dx \sqrt{-g}
\Bigl[- \frac{\Lambda}{8 \pi G}
+ \frac{R}{16 \pi G}
+ \frac{\alpha_1}{4 \pi} R^2 +
\frac{\alpha_2}{4 \pi} R_{\mu \nu} R^{\mu\nu}
+ \frac{\alpha_3}{4 \pi} R_{\alpha \beta \mu \nu}
R^{\alpha \beta \mu \nu} \Bigr]
\end{equation}
where $\Lambda$ is the cosmological constant,
$G$ the gravitational constant, and
$\alpha_i$  the coupling constants.
These constants are bare
ones that are to be renormalized.
The spacetime that we are particularly
interested in is a black hole background
minimally coupled to a massive scalar field.
The one-loop effective action of the scalar field
of a mass $m$
can be found by the DeWitt-Schwinger method~\cite{dewitt,birrel}
\begin{equation}
W_D (m) = \frac{1}{2 \bigl(4 \pi \bigr)^{D/2}}
\int d^Dx \sqrt{-g} \int_0^{\infty} d(is)
\sum_{k = 0}^{\infty} a_k (x)\bigl(is \bigr)^{k - 1 - D/2}
e^{- i m^2 s},
\end{equation}
where
\begin{equation}
a_0 = 1,~~
a_1 = \frac{1}{6} R,~~
a_2 = \frac{1}{30} R_{,\mu}^{~~;\mu} +
\frac{1}{72} R^2 + \frac{1}{180} R_{\alpha \beta \mu \nu}
R^{\alpha \beta \mu \nu} - \frac{1}{180} R_{\mu \nu}R^{\mu \nu}.
\end{equation}

The effective action $W_D$ involves divergent terms from the lower
limit $s=0$.
According to the Pauli-Villars regularization method,
we introduce a number of
bosonic and fermionic regulator fields of masses $m_{B_i}$
and $m_{F_i}$, respectively.
All the regulator and scalar fields contribute
to the effective action
\begin{equation}
W_D = \frac{1}{2 \bigl(4 \pi \bigr)^{D/2}}
\int d^Dx \sqrt{-g} \int_0^{\infty} d(is)
\sum_{k = 0}^{\infty} a_k (x)\bigl(is \bigr)^{k - 1 - D/2}
\bigl(\sum_i e^{- i m_{B_i}^2 s} - \sum_i e^{-i m_{F_i}^2 s}
\bigr).
\end{equation}

The effective action obtained by
the dimensional regularization method
shows different divergent structures
depending on even or odd dimensions.
In the Pauli-Villars regularization method
the divergent structure of the even dimensional case
differs from that of the odd dimensional case.

\subsection{Even Dimensions}

We first consider even dimensions.
Using the analytical continuation of
the integral (\ref{A10}) in the Appendix A,
we obtain the divergent contributions to
the effective action
in an even dimension $D=2n$:
\begin{eqnarray}
W^{{\rm div}}_{2n} = && \frac{1}{2 \bigl(4 \pi \bigr)^{n}}
\int d^{2n} x \sqrt{-g}
\sum_{k = 0}^{n} a_k (x) (-1)^{n + 1 - k}
\nonumber\\
&& \times
\Biggl[\frac{1}{(n-k)!} \Bigl(
\sum_i m_{B_i}^{2(n-k)}
\ln\bigl( m_{B_i}^2 \bigr) -
\sum_i m_{F_i}^{2(n-k)}
\ln\bigl( m_{F_i}^2 \bigr) \Bigr)
\nonumber\\
&& - \frac{1}{(n-k)!} \Bigl(I_1 +
\sum_{p = 1}^{n-k} \frac{1}{p} \Bigr)
\Bigl( \sum_i m_{B_i}^{2(n-k)}
- \sum_i m_{F_i}^{2(n-k)}
 \Bigr)
\nonumber\\
&& + \sum_{l = 2}^{n+1-k}
\frac{(-1)^l}{(n+1-k-l)!} I_{l}
\Bigl( \sum_i m_{B_i}^{2(n+1-k-l)}
- \sum_i m_{F_i}^{2(n+1-k-l)}
 \Bigr) \Biggr].
\end{eqnarray}
To remove
the infinite constants $I_p$ given by
Eq. (\ref{A11}),
we impose the mass conditions
\begin{equation}
\sum_i m_{B_i}^{2(n-k)} =
\sum_i m_{F_i}^{2(n-k)}
\label{ma}
\end{equation}
for $k = 0, 1, \cdots, n$.
We are then left with the renormalized
action
\begin{equation}
W^{{\rm ren}}_{2n} =
\int d^{2n} x \sqrt{-g}
\sum_{k = 0}^{n} a_k (x)
\frac{{\cal B}_k}{2 \bigl(4 \pi \bigr)^{n} (n-k)!}
\end{equation}
where
\begin{equation}
{\cal B}_k =
(-1)^{n + 1 - k}
\Bigl(
\sum_i m_{B_i}^{2(n-k)}
\ln\bigl( m_{B_i}^2 \bigr) -
\sum_i m_{F_i}^{2(n-k)}
\ln\bigl( m_{F_i}^2 \bigr) \Bigr)
\label{ren con1}
\end{equation}
are the renormalization constants.
We may now renormalize the one-loop
effective action for gravity by quantum matter field
by redefining the cosmological,
gravitational, and coupling constants
\begin{eqnarray}
&& \frac{\Lambda}{8\pi G}
- \frac{{\cal B}_0}{2 (4\pi)^n n!} =
\frac{\Lambda^{\rm ren}}{8\pi G^{\rm ren}},
\nonumber\\
&& \frac{1}{16\pi G}
+ \frac{{\cal B}_1}{12 (4\pi)^n (n-1)!} =
\frac{1}{16\pi G^{\rm ren}},
\nonumber\\
&& \frac{\alpha_1}{4\pi}
+ \frac{{\cal B}_2}{144 (4\pi)^n (n-2)!} =
\frac{\alpha^{\rm ren}_1}{4\pi},
\nonumber\\
&& \frac{\alpha_2}{4\pi}
- \frac{{\cal B}_2}{360 (4\pi)^n (n-2)!} =
\frac{\alpha^{\rm ren}_2}{4\pi},
\nonumber\\
&& \frac{\alpha_3}{4\pi}
+ \frac{{\cal B}_2}{360 (4\pi)^n (n-2)!} =
\frac{\alpha^{\rm ren}_3}{4\pi}.
\label{ren1}
\end{eqnarray}
For instance, in $D=4$  the conditions are
such that
the number of bosonic and fermionic fields
are equal
\begin{equation}
N_B = N_F =3
\end{equation}
and  the masses satisfy
\begin{eqnarray}
&& \sum_{i=1}^{3} m_{B_i}^{2}
= \sum_{i=1}^{3} m_{F_i}^{2},
\nonumber\\
&& \sum_{i=1}^{3} m_{B_i}^{4}
= \sum_{i=1}^{3} m_{F_i}^{4}.
\end{eqnarray}
One may choose a simple solution~\cite{demers}
\begin{eqnarray}
&& m_{B_1} = m,~ m_{B_2} = m_{B_3} = \sqrt{m^2 +
3 \mu^2},
\nonumber\\
&& m_{F_1} = m_{F_2} = \sqrt{m^2 + \mu^2},~
m_{F_3} = \sqrt{m^2 + 4 \mu^2}.
\end{eqnarray}
The renormalization constant related with the
gravitational constant can be written as
\begin{equation}
{\cal B}_1 =  m^2 \ln \Bigl(
\frac{m^2 (m^2 + 3 \mu^2)^2}{(m^2 + \mu^2)^2 (m^2 + 4 \mu^2)}
\Bigr)
+ \mu^2 \ln \Bigl(
\frac{(m^2 + 3 \mu^2)^6}{(m^2 + \mu^2)^2 (m^2 + 4 \mu^2)^4}
\Bigr),
\end{equation}
and becomes for a large $\mu$
\begin{equation}
{\cal B}_1 \simeq  m^2 \ln \Bigl(
\frac{3^2 m^2}{4 \mu^2}
\Bigr)
+ 2 \mu^2 \ln \Bigl(
\frac{3^3}{2^4} \Bigr).
\end{equation}
The other renormalization constant is
\begin{equation}
{\cal B}_2 =  - \ln \Bigl(
\frac{m^2 (m^2 + 3 \mu^2)^2}{(m^2 + \mu^2)^2 (m^2 + 4 \mu^2)}
\Bigr),
\end{equation}
which becomes for a large $\mu$
\begin{equation}
{\cal B}_2 \simeq  - \ln \Bigl(
\frac{3^2 m^2}{4 \mu^2}
\Bigr).
\end{equation}

In $D=6$, one has an additional condition
on the masses
\begin{equation}
\sum_i m_{B_i}^{6} = \sum_i m_{F_i}^{6},
\end{equation}
one of whose simple solutions is
\begin{eqnarray}
&& m_{B_1} = m,~ m_{B_2} = \sqrt{m^2 + 3 \mu^2},~
m_{B_3} = \sqrt{m^2 + 4 \mu^2},~
m_{B_4} = \sqrt{m^2 + 7 \mu^2},
\nonumber\\
&& m_{F_1} = m_{F_2} = \sqrt{m^2 + \mu^2},~
m_{F_3} = m_{F_4} = \sqrt{m^2 + 6 \mu^2}.
\end{eqnarray}
We rewrite ${\cal B}_1$ as
\begin{eqnarray}
{\cal B}_1 =  - m^4 \ln \Bigl(
\frac{m^2 (m^2 + 3 \mu^2) (m^2 + 4 \mu^2)
(m^2 + 7 \mu^2)}{(m^2 + \mu^2)^2 (m^2 + 6 \mu^2)^2}
\Bigr)
\nonumber\\
- 2m^2 \mu^2 \ln \Bigl(
\frac{(m^2 + 3 \mu^2)^3 (m^2 + 4 \mu^2)^4
(m^2 + 7 \mu^2)^7}{(m^2 + \mu^2)^2 (m^2 + 6 \mu^2)^12} \Bigr)
\nonumber\\
-  \mu^4 \ln \Bigl(
\frac{(m^2 + 3 \mu^2)^9 (m^2 + 4 \mu^2)^16
(m^2 + 7 \mu^2)^49}{(m^2 + \mu^2)^2 (m^2 + 6 \mu^2)^72} \Bigr).
\end{eqnarray}
For a large $\mu$ it becomes
\begin{equation}
{\cal B}_1 \simeq  - m^4 \ln \Bigl(
\frac{7 m^2}{3 \mu^2}
\Bigr) - 2 m^2 \mu^2 \ln \Bigl(
\frac{7^5}{2^4 3^9} \Bigr)
- \mu^4 \ln \Bigl(\frac{7^7}{2^{40} 3^{63}} \Bigr).
\end{equation}
The other two renormalization constants are
\begin{eqnarray}
{\cal B}_2 =   m^2 \ln \Bigl(
\frac{m^2 (m^2 + 3 \mu^2) (m^2 + 4 \mu^2)
(m^2 + 7 \mu^2)}{(m^2 + \mu^2)^2 (m^2 + 6 \mu^2)^2}
\Bigr)
\nonumber\\
+ \mu^2 \ln \Bigl(
\frac{(m^2 + 3 \mu^2)^3 (m^2 + 4 \mu^2)^4
(m^2 + 7 \mu^2)^7}{(m^2 + \mu^2)^2 (m^2 + 6 \mu^2)^12} \Bigr),
\end{eqnarray}
and
\begin{equation}
{\cal B}_3 =  -  \ln \Bigl(
\frac{m^2 (m^2 + 3 \mu^2) (m^2 + 4 \mu^2)
(m^2 + 7 \mu^2)}{(m^2 + \mu^2)^2 (m^2 + 6 \mu^2)^2}
\Bigr).
\end{equation}
It should be noted that
${\cal B}_1$, ${\cal B}_2$, and ${\cal B}_3$
can be regarded as independent
for an arbitrary large $\mu$, since they involve
different powers of $\mu$. That is, there is a unique
representation of $\ln (\mu^2)$, $\mu^2$,
and $\mu^4$ in terms of
${\cal B}_1$, ${\cal B}_2$, and ${\cal B}_3$.

\subsection{Odd Dimensions}

In odd dimensions
we use similarly the integral formula (\ref{A23})
in the Appendix A.
We then obtain the divergent
contributions to the effective action
in an odd dimension $D=2n+1$:
\begin{eqnarray}
W^{{\rm div}}_{2n+1} = && \frac{1}{2 \bigl(4 \pi \bigr)^{(2n+1)/2}}
\int d^{2n+1} x \sqrt{-g}
\sum_{k = 0}^{n} a_k (x) (-1)^{n + 1 - k}
\nonumber\\
&& \times
\Biggl[\frac{2^{2(n+1-k)}\pi^{1/2} (n+1-k)!}{(2n+2-2k)!}
\Bigl(
\sum_i m_{B_i}^{2n+1-2k}
\ln\bigl( m_{B_i}^2 \bigr) -
\sum_i m_{F_i}^{2n+1-2k}
\ln\bigl( m_{F_i}^2 \bigr) \Bigr)
\nonumber\\
&& + \sum_{l = 1}^{n+1-k}
\frac{(-1)^l}{(n+1-k-l)!} I_{(2l+1)/2}
\Bigl( \sum_i m_{B_i}^{2(n+1-k-l)}
- \sum_i m_{F_i}^{2(n+1-k-l)}
 \Bigr) \Biggr].
\end{eqnarray}
Imposing again the same mass conditions (\ref{ma})
as in $D=2n$,
we get the renormalized effective action
\begin{equation}
W^{{\rm ren}}_{2n+1} =
\int d^{2n} x \sqrt{-g}
\sum_{k = 0}^{n} a_k (x)
\frac{2^{2(n+1-k)} \pi^{1/2}(n+1-k)!
{\cal B}_k}{2 \bigl(4 \pi \bigr)^{(2n+1)/2} (2n+2-2k)!}
\end{equation}
where
\begin{equation}
{\cal B}_k =
(-1)^{n + 1 - k}
\Bigl(
\sum_i m_{B_i}^{2n+1-2k} -
\sum_i m_{F_i}^{2n+1-2k} \Bigr)
\label{ren con2}
\end{equation}
are the renormalization constants.
At one-loop level the cosmological,
gravitational, and coupling constants
can be renormalized by
\begin{eqnarray}
&& \frac{\Lambda}{8\pi G}
- \frac{(n+1)!{\cal B}_0}{\pi^{n} (2n+2)!}
= \frac{\Lambda^{\rm ren}}{8\pi G^{\rm ren}},
\nonumber\\
&& \frac{1}{16\pi G}
+
\frac{n!{\cal B}_1}{4 \pi^{n} (2n)!}
=
\frac{1}{16\pi G^{\rm ren}},
\nonumber\\
&& \frac{\alpha_1}{4\pi}
+
\frac{(n-1)!{\cal B}_2}{ 2^2 \cdot 144 \pi^{n} (2n-2)!}
=
\frac{\alpha^{\rm ren}_1}{4\pi},
\nonumber\\
&& \frac{\alpha_2}{4\pi}
-
\frac{(n-1)!{\cal B}_2}{ 2^3 \cdot 360 \pi^{n} (2n-2)!}
=
\frac{\alpha^{ren}_2}{4\pi},
\nonumber\\
&& \frac{\alpha_3}{4\pi}
+
\frac{(n-1)!{\cal B}_2}{ 2^3 \cdot 360 \pi^{n} (2n-2)!}
= \frac{\alpha^{\rm ren}_3}{4\pi}.
\label{ren2}
\end{eqnarray}

We have seen that the effective action
has
the different divergent structures depending on
the parity of dimension and
the renormalized effective action
of the even dimensions involves the different renormalization
constants from that of the odd dimensions.

\section{Number of States of Massive Scalar Field}

We now turn to the thermodynamic entropy of the
massive scalar field in a $D$-dimensional
Schwarzschild
and a nonextremal RN black hole.
But there seems to be no simple and systematic
method to evaluate the thermodynamic entropy
of a quantum field in an arbitrary dimensional
black hole background. We use directly the thermodynamic
definition of entropy
\begin{equation}
F = - \int_0^{\infty} dE \frac{g(E)}{e^{\beta E} - 1},
\label{fr en}
\end{equation}
where $g(E)$ is the number of states for a given $E$.
In order to find the free energy of a scalar field
it is necessary to find first the number of states.

The $D$-dimensional RN charged black hole
$(D \geq 4)$
has the metric \cite{myers}
\begin{equation}
ds^2 = - \Delta (r) dt^2 + \Delta^{-1} (r) dr^2
+ r^2 d \Omega_{D- 2}^2,
\label{n bh}
\end{equation}
where
\begin{equation}
\Delta (r) =
\Biggl(1 - \Bigl( \frac{r_-}{r}
\Bigr)^{D-3} \Biggr)
\Biggl(1 - \Bigl( \frac{r_+}{r}
\Bigr)^{D-3} \Biggr),
\end{equation}
where $r_+$ and $r_-$ are the outer
and inner event horizons given by
\begin{equation}
r_{\pm} =  \Biggl[
\frac{4 \Gamma
\Bigl(\frac{D -1}{2} \Bigr)}{ (D-2)
\pi^{(D-3)/2}}
\Bigl( M \pm \sqrt{M^2 - Q^2}
\Bigr) \Biggr]^{\frac{1}{D-3}},
\end{equation}
where $\Gamma$ is the gamma function.
The Schwarzschild black hole
is recovered
as a limiting case of
$Q = 0$.

The number of states~\cite{mann} of a massive scalar field
in the black hole background (\ref{n bh}) is given by
\begin{equation}
g_D(E,m) = \frac{1}{\pi} \int_{r_+ + h}^{L}
\frac{dr}{\Delta(r)} \int dl (2l+D-3) L_D (l)
\sqrt{E^2 - \Bigl(m^2 + \frac{l(l+D-3)}{r^2}
\Bigr) \Delta(r)},
\label{num st}
\end{equation}
where
\begin{equation}
L_D (l) = \frac{\Gamma(l+D-3)}{\Gamma(D-2)\Gamma(l+1)}
\label{ag}
\end{equation}
is a multiplication factor of
the degeneracy of angular momentum states.
We note that the number of states in this form can be
extended to a non-integral dimension.
By changing the variable
\begin{equation}
y = l(l+D-3),
\end{equation}
we rewrite the number of states as
\begin{equation}
g_D(E,m) = \frac{1}{\pi} \int_{r_+ + h}^{L}
\frac{dr}{r \sqrt{\Delta (r)}}
\int_0^{y_+} dy
L_D (l(y))
\sqrt{y_+ - y},
\label{y int}
\end{equation}
where the integration is restricted to
\begin{equation}
y_+ = (E^2 - m^2 \Delta (r)) \frac{r^2}{\Delta (r)}.
\end{equation}
We compute the number of states separately
in even and odd dimensions.

\subsection{Even Dimensions}

The multiplication factor of
the degeneracy of angular momentum states
in four dimensions is simply given by
\begin{equation}
L_4 = 1,
\end{equation}
and in $D=2n, (n \geq 3)$ by
\begin{eqnarray}
L_{2n} (y) &=& \frac{1}{\Gamma(2n-2)}
\prod_{k = 2}^{n-1} \Bigl(y
+ (2n-k-2)(k-1) \Bigr)
\nonumber\\
&\equiv& \frac{1}{\Gamma(2n-2)}
\sum_{k = 0}^{n-2} C^{2n}_k y^k,
\end{eqnarray}
whose coefficients are
\begin{eqnarray}
C^{2n}_{n-2} = && 1,
\nonumber\\
C^{2n}_{n-3} = && \frac{1}{3} (n-2) (n-1) (2n-3),
\nonumber\\
\vdots
\nonumber\\
C^{2n}_0 = && \prod_{k=2}^{n-2} (2n-k-2)(k-1).
\end{eqnarray}

The $y$-integration~\cite{grad} yields
the number of states
\begin{equation}
g_{2n}(E,m) = \frac{1}{\pi \Gamma(2n-2)}
\sum_{k=0}^{n-2} C_k^{2n} B\Bigl(k+1,
\frac{3}{2} \Bigr)
\int_{r_+ + h}^{L}
\frac{dr}{r \sqrt{\Delta (r)}} y_+^{k+ \frac{3}{2}},
\end{equation}
where $B$ is the beta function.

With the change of the variable
\begin{eqnarray}
x = && 1 - \Bigl(\frac{r_+}{r} \Bigr)^{2n-3},
\nonumber\\
\epsilon = && (2n-3)\frac{h}{r_+},
\end{eqnarray}
we obtain the number of states
\begin{eqnarray}
g_{2n} (E,m) = && \frac{1}{\pi \Gamma(2n-2)}
\sum_{k = 0}^{n-2}  C_k^{2n}
B \Bigl(k+1, \frac{3}{2} \Bigr)
\frac{r_+^{2k+3}}{2n-3}
\nonumber\\
&& \times
\int_{\epsilon} dx \frac{ \Bigl[E^2 - x(1-u+ux)m^2
\Bigr]^{k + \frac{3}{2}}}{(1-x)^{1+\frac{2k+3}{2n-3}}
x^{k+2}(1-u+ux)^{k+2}},
\label{f ns}
\end{eqnarray}
where $u = \Bigl(\frac{r_-}{r_+} \Bigr)^{2n-3}$.

The most interesting ultraviolet
divergences of free energy and entropy  come from
the part of the number of states that are the divergent
functions of the brick wall thickness
near the event horizon.
So focussing on the lower limit of the integration,
we expand the denominator around $x = 0$
\begin{equation}
\frac{1}{(1 -x)^{1 + \frac{2k + 3}{2n-3}}
(1 - u + ux)^{k+ 2}} = \sum_{q=0}^{\infty}
H_q^{2n,k} x^q,
\end{equation}
where $H_q^{2n,k}$ are the coefficients of
Taylor expansion, 
and substitute into Eq. (\ref{f ns}).
To extract the divergent parts of Eq. (\ref{f ns}),
we transform the quadratic
\begin{equation}
E^2 - x(1 - u + ux)m^2 =
 \Bigl(\frac{(1-u)^2 m^2 + 2uE^2}{
(1-u)^2 m^2} \Bigr) Z
\Bigl(1 - \frac{u}{(1-u)^2 m^2 + 2uE^2}
\bigl( Z + \frac{E^4}{Z} \bigr) \Bigr)
\end{equation}
where
\begin{equation}
Z = E^2 - (1 - u) m^2 x,
\end{equation}
and  expand the numerator
\begin{eqnarray}
\bigl( E^2 - x(1 - u + ux)m^2 \bigr)^{k
+ \frac{3}{2}} = &&
  \Bigl(\frac{(1-u)^2 m^2 + 2uE^2}{
(1-u)^2 m^2} \Bigr)^{k+ \frac{3}{2}}
Z^{k + \frac{3}{2}}
\nonumber\\
&& \times
\sum_{p = 0}^{\infty} (-1)^p
{{k+ \frac{3}{2}} \choose {p}}
\Bigl(
\frac{u}{(1-u)^2 m^2 + 2uE^2}\Bigr)^{p}
\nonumber\\
&& \times
\sum_{l = 0}^{p} {p \choose l}
E^{4l} Z^{p -2l}
\end{eqnarray}
Finally we get the number of states
\begin{eqnarray}
g_{2n} (E,m) = && \frac{1}{\pi \Gamma(2n-2)}
\sum_{k = 0}^{n-2}  C_k^{2n}
B \Bigl(k+1, \frac{3}{2} \Bigr)
\frac{r_+^{2k+3}}{2n-3}
\sum_{q = 0}^{\infty} H_q^{2n,k}
\sum_{p = 0}^{\infty} (-1)^p
{{k+ \frac{3}{2}} \choose {p}}
\nonumber\\
&& \times
\Bigl(\frac{(1-u)^2 m^2 + 2uE^2}{
(1-u)^2 m^2} \Bigr)^{k+ \frac{3}{2}}
\Bigl(
\frac{u}{(1-u)^2 m^2 + 2uE^2}\Bigr)^{p}
\nonumber\\
&& \times
\sum_{l = 0}^{p} {p \choose l}
E^{4l}
\int_{\epsilon} dx
x^{-k-2 + q}
Z^{k +p -2l + \frac{3}{2}},
\end{eqnarray}
whose integrations can be done relatively easily.

From the integral formulas in the Appendix B,
we see that the $x$-integrals for $k+2 < q$
yield terms of the order of
$O\bigl(\frac{1}{m^2}\bigr)$,
and for $k+2 > q$ and $k+p-2l + \frac{3}{2} < 0$
also lead to terms of the order of
$O\bigl(\frac{1}{m^2} \bigr)$
even though they involves ultraviolet divergent
factors. Therefore all these integrals can be
neglected in the large mass limit.
The only divergent and nonvanishing terms come from
the integrals with $k+2 >q$ and $k+p - 2l
+ \frac{3}{2} >0$. These integrals lead to the
divergent structure
\begin{equation}
E^{2k + 2p - 6l + 3} m^{2l}
\frac{1}{\epsilon^{k+1-q -l}}, ~ l = 0, 1, \cdots, k - q,
\end{equation}
and
\begin{equation}
E^{2p + 2q - 4l + 1} m^{2k+2-2q} \ln (\epsilon).
\end{equation}
We find the most divergent term of the number of states
\begin{equation}
g_{2n}^{\rm m. div} = \frac{1}{(n-1)(2n-3) \pi \Gamma(2n-2)}
B \bigl(n-1, \frac{3}{2} \bigr)
r_+^{2n-1} \Bigl(\frac{(1-u)^2 m^2 + 2uE^2}{
(1-u)^2 m^2} \Bigr)^{n- \frac{1}{2}}
\frac{E^{2n-1}}{\epsilon^{n-1}}.
\end{equation}
The most divergent term is used to give rise to
the Bekenstein-Hawking entropy in the brick wall
model ~\cite{mann}. The most divergent term
differs from that in Ref.~\cite{mann}.
Moreover, it should be noted that the most divergent term
is to be removed in the Pauli-Villars regularization method
as will be shown in the next sections.

\subsection{Odd Dimensions}

We repeat the computation of the number
of states in odd dimensions, which differs slightly
from that in even dimensions.
First, in five dimensions
we have a quite
simple multiplication factor of the angular momentum degeneracy
\begin{equation}
L_5 = \frac{1}{2} \sqrt{y+1}.
\label{5an}
\end{equation}
In a general odd dimension $D=2n+1, (n \geq 3)$
the number of states takes the form
\begin{eqnarray}
L_{2n+1} (y) &=&  \frac{1}{\Gamma(2n-1)}
\sqrt{ y + (n-1)^2}
\prod_{k = 2}^{n-1} \Bigl(y
+ (2n-k-1)(k-1) \Bigr)
\nonumber\\
&\equiv& \frac{1}{\Gamma(2n-1)} \sum_{k = 0}^{n-2} C^{2n+1}_k y^k
\sqrt{ y + (n-1)^2},
\end{eqnarray}
whose coefficients are
\begin{eqnarray}
C^{2n+1}_{n-2} = && 1,
\nonumber\\
C^{2n+1}_{n-3} = && \frac{1}{6} (n-2) (n-1) (4n-3),
\nonumber\\
\vdots
\nonumber\\
C^{2n+1}_0 = && \prod_{k=2}^{n-2} (2n-k-1)(k-1).
\end{eqnarray}
Note that the multiplication factor of the degeneracy of
angular momentum states of odd dimensions
has an additional factor of the square root
which differs from that of even dimensions.

To get the number of states
we do the intermediate $y$-integration
(\ref{y int}).
The $y$-integration can be done
recursively
\begin{eqnarray}
J_k &=& \int_0^{y_+} dy y^k \sqrt{
\bigl( y + (n-1)^2
\bigr) \bigl( y_+ -y \bigr)}
\nonumber\\
&=& \frac{2k+1}{2(k+2)}
\Bigl(y_+ - (n-1)^2 \Bigr)
J_{k-1} + \frac{k-1}{k+2} (n-1)^2
y_+ J_{k-2},
\label{rec re}
\end{eqnarray}
whose two lowest integrals are explicitly
\begin{eqnarray}
J_1 = && \frac{(n-1)^3}{3} y_+^{\frac{3}{2}}
+ \frac{n-1}{8}
\Bigl(y_+ - (n-1)^2 \Bigr)^2
y_+^{\frac{1}{2}}
\nonumber\\
&& - \frac{1}{8}
\Bigl(y_+ - (n-1)^2 \Bigr)
\Bigl(y_+ + (n-1)^2 \Bigr)^2
\Biggl(\frac{\pi}{2} +
\arcsin \Bigl(\frac{y_+ - (n-1)^2}{y_+ + (n-1)^2} \Bigr) \Biggr),
\nonumber\\
J_0 = &&
 \frac{n-1}{4}
\Bigl(y_+ - (n-1)^2 \Bigr)
y_+^{\frac{1}{2}}
+ \frac{1}{8}
\Bigl(y_+ + (n-1)^2 \Bigr)^2
\Biggl(\frac{\pi}{2} +
\arcsin \Bigl(\frac{y_+ -(n-1)^2}{y_+ + (n-1)^2} \Bigr) \Biggr).
\end{eqnarray}
Near the event horizon ($y_+ >> 1$)
they are further approximated
by
\begin{eqnarray}
J_1 = && \frac{(n-1)^3}{3} y_+^{\frac{3}{2}}
+ \frac{n-1}{8}
\Bigl(y_+ - (n-1)^2 \Bigr)^2
y_+^{\frac{1}{2}}
\nonumber\\
&& - \frac{1}{8}
\Bigl(y_+ - (n-1)^2 \Bigr)
\Bigl(y_+ + (n-1)^2 \Bigr)^2
\Biggl(\pi - 2 (n-1)^2
\frac{1}{y_+^{\frac{1}{2}}}
\Biggr),
\nonumber\\
J_0 = &&
 \frac{n-1}{4}
\Bigl(y_+ - (n-1)^2 \Bigr)
y_+^{\frac{1}{2}}
+ \frac{1}{8}
\Bigl(y_+ + (n-1)^2 \Bigr)^2
\Biggl(\pi - 2 (n-1)^2
\frac{1}{y_+^{\frac{1}{2}}}
\Biggr).
\end{eqnarray}
We can rearrange
the number of states
\begin{equation}
g_{2n+1}(E,m) = \frac{1}{\pi \Gamma(2n-1)}
\sum_{k=0}^{n-2} C_k^{2n+1}
\int_{r_+ + h}^{L}
\frac{dr}{r \sqrt{\Delta (r)}} J_k (y_+),
\end{equation}
as
\begin{equation}
g_{2n+1}(E,m) = \frac{1}{\pi \Gamma(2n-1)}
\sum_{k=-1}^{2n} F_k^{2n+1}
\int_{r_+ + h}^{L}
\frac{dr}{r \sqrt{\Delta (r)}} y_+^{\frac{k}{2}},
\end{equation}
where $F_k^{2n+1}$ is the coefficient of $y_+^{\frac{k}{2}}$
which is determined by solving recursive relations
(\ref{rec re}) as a power series of $y_+$.
The number of states in odd dimensions
differs from that in even dimensions
by integral powers of $y_+$.

Similarly as in the even dimensional case
we change the variable
\begin{equation}
x = 1 - \Bigl(\frac{r_+}{r} \Bigr)^{2n-2},
\end{equation}
and replace a brick wall $h$
by $\epsilon = (2n-2)\frac{h}{r_+}$.
The radial integration of the
number of states becomes
\begin{eqnarray}
g_{2n+1} (E,m) = && \frac{1}{\pi \Gamma(2n-1)}
\sum_{k = -1}^{2n} F_k^{2n+1}
\frac{r_+^{k}}{2n-2}
\nonumber\\
&& \times
\int_{\epsilon} dx \frac{ \Bigl[E^2 - x(1-u+ux)m^2
\Bigr]^{\frac{k}{2}}}{(1-x)^{1+\frac{k}{2n-2}}
x^{\frac{k+1}{2}}(1-u+ux)^{\frac{k+1}{2}}},
\label{f ns2}
\end{eqnarray}
where $u = \Bigl(\frac{r_-}{r_+} \Bigr)^{2n-2}$.

Expanding the denominator around $x = 0$
\begin{equation}
\frac{1}{(1 -x)^{1 + \frac{k}{2n-2}}
(1 - u + ux)^{\frac{k+1}{2}}} = \sum_{q=0}^{\infty}
K_q^{2n+1,k} x^q,
\end{equation}
where $K_q^{2n+1,k}$ are the coefficients of
Taylor expansion, and the numerator
\begin{eqnarray}
\bigl( E^2 - x(1 - u + ux)m^2 \bigr)^{
\frac{k}{2}} = &&
  \Bigl(\frac{(1-u)^2 m^2 + 2uE^2}{
(1-u)^2 m^2} \Bigr)^{ \frac{k}{2}}
Z^{\frac{k}{2}}
\nonumber\\
&& \times
\sum_{p = 0}^{\infty} (-1)^p
{{\frac{k}{2}} \choose {p}}
\Bigl(
\frac{u}{(1-u)^2 m^2 + 2uE^2}\Bigr)^{p}
\nonumber\\
&& \times
\sum_{l = 0}^{p} {p \choose l}
E^{4l} Z^{p -2l},
\end{eqnarray}
we finally get the number of states
\begin{eqnarray}
g_{2n+1} (E,m) = && \frac{1}{\pi \Gamma(2n-1)}
\sum_{k = -1}^{2n}  F_k^{2n+1}
\frac{r_+^{k}}{2n-2}
\sum_{q = 0}^{\infty} K_q^{2n+1,k}
\sum_{p = 0}^{\infty} (-1)^p
{{\frac{k}{2}} \choose {p}}
\nonumber\\
&& \times
\Bigl(\frac{(1-u)^2 m^2 + 2uE^2}{
(1-u)^2 m^2} \Bigr)^{\frac{k}{2}}
\Bigl(
\frac{u}{(1-u)^2 m^2 + 2uE^2}\Bigr)^{p}
\nonumber\\
&& \times
\sum_{l = 0}^{p} {p \choose l}
E^{4l}
\int_{\epsilon} dx
x^{- \frac{k+1}{2} + q}
Z^{\frac{k}{2} + p -2l}.
\end{eqnarray}

We may find the divergent structure of the
number of states
by doing directly the polynomial integrals for
even integers $k$ and computing the same integrals
for odd integers $k$ as in the even dimensional case.
It is not difficult to find
the most divergent term
\begin{equation}
g_{2n+1}^{\rm m. div} = \frac{1}{2^{2n-1} \bigl(n- \frac{1}{2}
\bigr)(2n-2) (n-1)!
n!}
r_+^{2n} \Bigl(\frac{(1-u)^2 m^2 + 2uE^2}{
(1-u)^2 m^2} \Bigr)^{n}
\frac{E^{2n}}{\epsilon^{n-\frac{1}{2}}}.
\end{equation}

\section{Free Energy in
the Five- and Six-Dimensional RN Black Holes}

We work out explicitly the free energy and
thermodynamic entropy
of the scalar field in the
five- and six-dimensional RN black hole
backgrounds.  These examples are enough to show
the  difference between even and odd dimensionality.

\subsection{Five-Dimensional RN Black Hole}

We find explicitly the number of states in the five-dimensional
RN black hole. Substituting (\ref{5an}) into (\ref{y int}),
the exact number of states  is found
\begin{equation}
g_5 (E,m) = \frac{1}{16\pi}
\int_{r_+ + h}^L \frac{dr}{r \sqrt{\Delta (r)}}
\Biggl(2 \bigl(y_+ -1 \bigr)
\sqrt{y_+} +  \bigl(y_+ +1
\bigr)^2 \Bigl( \frac{\pi}{2}
+ \arcsin \Bigl( \frac{y_+ -1}{y_+ +1}\Bigr)
\Bigr)\Biggr).
\end{equation}
Near the event horizon $y_+$ becomes large
and we may approximate the number of states as
\begin{equation}
g_5 (E,m) = \frac{1}{16}
\int_{r_+ + h}^L \frac{dr}{r \sqrt{\Delta (r)}}
\Bigl(y_+^2
+ 2 y_+ - \frac{6}{\pi} y_+^{\frac{1}{2}} \Bigr).
\end{equation}
With the change of variable used in Sect. III,
we rewrite the number of states as
\begin{eqnarray}
g_5 (E,m) = &&
\frac{r_+^4}{32}
\int_{\epsilon} dx
\frac{\Bigl[E^2 - x (1-u+ux)m^2 \Bigr]^{2}}{
x^{\frac{5}{2}} (1-u+ux)^{\frac{5}{2}} (1-x)^{3}}
\nonumber\\
 + &&
\frac{r_+^2}{16}
\int_{\epsilon} dx
\frac{\Bigl[E^2 - x (1-u+ux)m^2 \Bigr]}{
x^{\frac{3}{2}} (1-u+ux)^{\frac{3}{2}} (1-x)^{2}}
\nonumber\\
- &&
\frac{3r_+}{16 \pi}
\int_{\epsilon} dx
\frac{\Bigl[E^2 - x (1-u+ux)m^2 \Bigr]^{\frac{1}{2}}}{
x (1-u+ux) (1-x)^{\frac{3}{2}}}.
\end{eqnarray}
After expanding the denominators of the integrals
around $x=0$, doing integrals, and taking the limit of
large mass, we obtain the number of
states
\begin{eqnarray}
g_5 (E,m) = &&
\frac{r_+^4}{32}
\Biggl[\frac{2}{3(1-u)^{\frac{5}{2}}}
\frac{E^4}{\epsilon^{\frac{3}{2}}}
+ \Bigl(\frac{6-11u}{(1-u)^{\frac{7}{2}}} E^2
- \frac{4}{(1-u)^{\frac{3}{2}}} m^2  \Bigr)
\frac{E^2}{\epsilon^{\frac{1}{2}}}
+ O \bigl(\epsilon^{\frac{1}{2}} \bigr) \Biggr]
\nonumber\\
+ && \frac{r_+^2}{16} \Biggl[\frac{2}{(1-u)^{\frac{3}{2}}}
\frac{E^4}{\epsilon^{\frac{1}{2}}}
+ O \bigl(\epsilon^{\frac{1}{2}} \bigr) \Biggr]
\nonumber\\
- &&
\frac{3r_+}{16 \pi}
\Biggl[
\frac{1}{(1-u)} A^1_{\frac{1}{2}}
- \frac{u}{2 (1-u)^3 m^2} A^1_{\frac{3}{2}}
+ O \bigl(\epsilon \bigr)
+ O \bigl( \frac{1}{m} \bigr)
\Biggr],
\end{eqnarray}
where $A^{q}_{\frac{2p-1}{2}}$ are integrals
defined in Appendix B.
The ultraviolet divergent part of $g_5$ is
\begin{eqnarray}
g_5^{{\rm div}} (E,m) = &&
\frac{r_+^4}{32}
\Biggl[\frac{2}{3(1-u)^{\frac{5}{2}}}
\frac{E^4}{\epsilon^{\frac{3}{2}}}
+ \Bigl(\frac{6-11u}{(1-u)^{\frac{7}{2}}} E^2
- \frac{4}{(1-u)^{\frac{3}{2}}} m^2  \Bigr)
\frac{E^2}{\epsilon^{\frac{1}{2}}} \Biggr]
\nonumber\\
+ && \frac{r_+^2}{16} \Biggl[\frac{2}{(1-u)^{\frac{3}{2}}}
\frac{E^4}{\epsilon^{\frac{1}{2}}} \Biggr]
+
\frac{3r_+}{16 \pi}
\Biggl[
\frac{1}{(1-u)} E \ln (\epsilon) \Biggr].
\label{div5}
\end{eqnarray}
The remaining part to be renormalized
in the large mass limit is
\begin{eqnarray}
g^{{\rm ren}}_5 (E,m)
=
 \frac{3r_+}{16 \pi}
\Biggl[\frac{2}{(1-u)}  E
+ \frac{1}{(1-u)} E \ln \Bigl( \frac{(1-u)m^2}{4E^2}
\Bigr) \Biggr].
\end{eqnarray}

From the definition (\ref{fr en}), the free
energy consists of two parts,
\begin{equation}
F_5 (m) = F^{{\rm div}}_5 (m)
+ F^{{\rm ren}}_5 (m),
\label{fr5}
\end{equation}
where the ultraviolet
divergent part is
\begin{eqnarray}
F_5^{{\rm div}} (m) =
- && \frac{r_+^4}{32}
\Biggl[\frac{2\Gamma(5)\zeta(6)}{3(1-u)^{\frac{5}{2}}
\beta^5}
\frac{1}{\epsilon^{\frac{3}{2}}}
+  \Bigl(\frac{(6-11u)\Gamma(5)\zeta(5)}{(1-u)^{\frac{7}{2}}
\beta^4}
- \frac{4 \Gamma(3)\zeta(3)}{(1-u)^{\frac{3}{2}}
\beta^3} m^2  \Bigr)
\frac{1}{\epsilon^{\frac{1}{2}}} \Biggr]
\nonumber\\
- && \frac{r_+^2}{16} \Biggl[\frac{2 \Gamma(5)
\zeta(5) }{(1-u)^{\frac{3}{2}} \beta^5}
\frac{1}{\epsilon^{\frac{1}{2}}} \Biggr]
-
\frac{3r_+}{16 \pi}
\Biggl[
\frac{\Gamma(2) \zeta(2) }{(1-u) \beta^2}
 \ln (\epsilon) \Biggr],
\end{eqnarray}
where $\zeta$ is the
Riemann zeta function,
and the part to be renormalized
in the large mass limit is
\begin{equation}
F^{{\rm ren}}_5 (m)
=
-  \frac{5r_+}{16 \pi}
\Biggl[\frac{2 \Gamma(2) \zeta(2)}{(1-u) \beta^2}
+ \frac{\Gamma(2) \zeta(2)}{(1-u) \beta^2}
\ln \Bigl( \frac{(1-u)m^2}{4 \beta^2}
\Bigr)
- \frac{\psi_{1}}{(1-u) \beta^2}
\Biggr],
\end{equation}
where
\begin{equation}
\psi_{k} = 2 \int_0^{\infty}
dt \frac{t^k \ln (t)}{e^t - 1}.
\end{equation}

\subsection{Six-Dimensional RN Black Hole}

From Sect. III we find the number of states
in the six dimensional RN black hole.
After integrating the angular momentum states,
the number of states is
\begin{eqnarray}
g_6 (E,m) =
\frac{2r_+^5}{135 \pi}
\int_{\epsilon} dx
\frac{\Bigl[E^2 - x (1-u+ux)m^2 \Bigr]^{\frac{5}{2}}}{
x^3 (1-u+ux)^3 (1-x)^{\frac{8}{3}}}
\nonumber\\
+
\frac{2r_+^3}{27 \pi}
\int_{\epsilon} dx
\frac{\Bigl[E^2 - x (1-u+ux)m^2 \Bigr]^{\frac{3}{2}}}{
x^2 (1-u+ux)^2 (1-x)^{2}}.
\end{eqnarray}
Doing $x$-integration explicitly and keeping
only divergent and nonvanishing terms for large
masses $ ( m >> 1)$ of regulator fields, we obtain
\begin{eqnarray}
g_6 (E,m) =
\frac{2r_+^5}{135 \pi}
\Biggl[
\Bigl(
\frac{1}{(1-u)^3}
A^3_{\frac{5}{2}}
+ \frac{8-17u}{3(1-u)^4}
A^2_{\frac{5}{2}}
+ \frac{44-160u+170u^2}{9(1-u)^5}
A^1_{\frac{5}{2}}
\Bigr)
\nonumber\\
-  \frac{5u}{2(1-u)^2 m^2}
\Bigl(
\frac{1}{(1-u)^3} \bigl(
A^3_{\frac{7}{2}} + E^4 A^3_{\frac{3}{2}} \bigr)
+
\frac{8-17u}{3(1-u)^4} \bigl(
A^2_{\frac{7}{2}} + E^4 A^2_{\frac{3}{2}} \bigr)
\Bigr)
\nonumber\\
+  \frac{15u^2}{4(1-u)^7 m^4}
\Bigl(
A^3_{\frac{9}{2}} + 2 E^4 A^3_{\frac{5}{2}}
+ E^8 A^3_{\frac{1}{2}}  \Bigr)
+ O \bigl(\frac{1}{m} \bigr) \Biggr]
\nonumber\\
+
\frac{2r_+^3}{27 \pi}
\Biggl[
\Bigl(
\frac{1}{(1-u)^2}
A^2_{\frac{3}{2}}
+ \frac{2-4u}{(1-u)^3}
A^1_{\frac{3}{2}} \Bigr)
- \frac{3u}{2(1-u)^4 m^2}
\Bigl(
A^2_{\frac{5}{2}} + E^4 A^2_{\frac{5}{2}}  \Bigr)
+ O\bigl(\frac{1}{m}\bigr) \Biggr]
\end{eqnarray}
where $A^{q}_{\frac{2p-1}{2}}$  are integrals
defined in the Appendix  B.
We suppressed the terms that vanish as
$m$ goes to infinity.
The number of states consists of the ultraviolet
divergence terms
\begin{eqnarray}
g^{{\rm div}}_6 (E,m) =   &&
\frac{2r_+^5}{135 \pi}
\Biggl[
\frac{1}{2(1-u)^3}
\frac{E^5}{\epsilon^2}
+ \frac{32-23u}{12(1-u)^4}
\frac{E^5}{\epsilon}
\nonumber\\
+ && \Bigl(
\frac{95u}{8(1-u)^3} m^2
- \frac{3457 - 1120u-2740u^2}{72(1-u)^5} E^2
\Bigr) E^3 \ln (\epsilon) \Biggr]
\nonumber\\
+ &&
\frac{2r_+^3}{27 \pi}
\Biggl[
\frac{1}{(1-u)^2}
\frac{E^3}{\epsilon}
- \frac{4 + u - 9u^2}{2(1-u)^3} E^3 \ln(\epsilon)
\Biggr]
\end{eqnarray}
and the remaining terms
\begin{eqnarray}
g^{{\rm ren}}_6 (E,m) =
- \frac{2r_+^5}{135 \pi}
\Biggl[
\frac{23}{4(1-u)}m^4 E
- \frac{368-547u}{6(1-u)^3}
m^2 E^3
+ \frac{16192 + 54400u-109715u^2}{1080 (1-u)^5}
E^5
\nonumber\\
-  \Bigl( \frac{15}{8(1-u)}m^4 E
- \frac{160-55u}{24(1-u)^3} m^2 E^3 +
\frac{352+1120u-635u^2}{72(1-u)^5} E^5
\Bigr) \ln \Bigl(\frac{(1-u)m^2}{4E^2} \Bigr) \Biggr]
\nonumber\\
+
\frac{2r_+^3}{27 \pi}
\Biggl[
\frac{4}{(1-u)} m^2 E
- \frac{16+7u}{3} E^3
+ \Bigl( \frac{3}{2(1-u)}m^2 E
- \frac{4 + u}{2(1-u)^3} E^3 \Bigr)
\ln\Bigl(\frac{(1-u)m^2}{4E^2} \Bigr)
\Biggr].
\end{eqnarray}

Likewise, the free energy consists of two parts
\begin{equation}
F_6 (m) = F^{{\rm div}}_6 +
F^{{\rm ren}}_6,
\label{fr6}
\end{equation}
where  the divergent part is
\begin{eqnarray}
F^{{\rm div}}_6 (m) =  &&
- \frac{2r_+^5}{135 \pi}
\Biggl[
\frac{\Gamma(6) \zeta(6)}{2(1-u)^3
\beta^6}
\frac{1}{\epsilon^2}
+ \frac{(32-23u) \Gamma(6) \zeta(6)}{12(1-u)^4
\beta^6}
\frac{1}{\epsilon}
\nonumber\\
+ && \Bigl(
\frac{95u \Gamma(4) \zeta(4) }{8(1-u)^3 \beta^4} m^2
- \frac{(3457 - 1120u-2740u^2)
\Gamma(6) \zeta(6)}{72(1-u)^5 \beta^6}
\Bigr)  \ln (\epsilon) \Biggr]
\nonumber\\
- &&
\frac{2r_+^3}{27 \pi}
\Biggl[
\frac{\Gamma(4) \zeta(4)}{(1-u)^2 \beta^4}
\frac{1}{\epsilon}
- \frac{(4 + u - 9u^2) \Gamma(4) \zeta (4)}{2(1-u)^3
\beta^4}  \ln(\epsilon)
\Biggr]
\end{eqnarray}
and the part to be renormalized is
\begin{eqnarray}
F^{{\rm ren}}_6 (m) =  &&
 \frac{2r_+^5}{135 \pi}
\Biggl[
\frac{23 \Gamma(2) \zeta(2)}{4(1-u) \beta^2}m^4
- \frac{(368-547u) \Gamma(4) \zeta(4)}{6(1-u)^3 \beta^4}
m^2
\nonumber\\
+ && \frac{(16192 + 54400u-109715u^2)
\Gamma(6) \zeta(6)}{1080 (1-u)^5 \beta^6}
\nonumber\\
+ && \Bigl( \frac{15 \Gamma(2) \zeta(2)}{8(1-u) \beta^2}m^4
- \frac{(160-55u) \Gamma(4) \zeta(4) }{24(1-u)^3 \beta^4} m^2
\nonumber\\
+ &&
\frac{(352+1120u-635u^2)
\Gamma(6) \zeta(6)}{72(1-u)^5 \beta^6}
\Bigr) \ln \Bigl(\frac{(1-u)m^2}{4\beta^2} \Bigr)
\nonumber\\
+ && \frac{(16192 + 54400u-109715u^2)
\Gamma(6) \zeta(6)}{1080 (1-u)^5 \beta^6}
\nonumber\\
- && \Bigl( \frac{15 \psi_{1}}{8(1-u) \beta^2}m^4
- \frac{(160-55u) \psi_{3} }{24(1-u)^3 \beta^4} m^2
+
\frac{(352+1120u-635u^2)
\psi_{5}}{72(1-u)^5 \beta^6}
\Bigr)
\Biggr]
\nonumber\\
- &&
\frac{2r_+^3}{27 \pi}
\Biggl[
\frac{4 \Gamma(2) \zeta(2)}{(1-u) \beta^2} m^2
- \frac{(16+7u) \Gamma(4) \zeta(4)}{3 \beta^4}
\nonumber\\
+ && \Bigl( \frac{3 \Gamma(2) \zeta(2)}{2(1-u) \beta^2}m^2
- \frac{(4 + u) \Gamma(4) \zeta(4)}{2(1-u)^3 \beta^4}  \Bigr)
\ln\Bigl(\frac{(1-u)m^2}{4\beta^2} \Bigr)
\nonumber\\
- && \Bigl( \frac{3 \psi_{1}}{2(1-u) \beta^2}m^2
- \frac{(4 + u) \psi_{3}}{2(1-u)^3 \beta^4}  \Bigr)
\Biggr].
\label{ren6}
\end{eqnarray}

\section{Pauli-Villars Regularization}

We now regularize the free energy using
the Pauli-Villars regularization method used
in Sect. II.
We assume that
the bosonic or fermionic regulator fields obey
the same Bose-Einstein distribution~\cite{demers}
\begin{equation}
F_D (m) = \mp \int_0^{\infty}
dE \frac{g_D (E,m)}{e^{\beta E} - 1},
\end{equation}
where the upper sign is for the bosonic fields
and the lower sign for the fermionic fields.

The main idea of the Pauli-Villars regularization is
to subtract the ultraviolet divergences of the
original scalar field by those of regulator fields.
To be more concrete we apply the Pauli-Villars
regularization method to the thermodynamic entropy
of the scalar field in five- and six-dimensional RN
black hole backgrounds.
We introduce a number of bosonic and fermionic
regulator fields  with masses $m_{B_i}$ and
$m_{F_i}$, respectively,
whose number and mass conditions
will imposed later such that all the ultraviolet
divergences and other unnecessary infinite quantities
disappear.
As mentioned earlier, the free energy of the fermionic
regulator fields has the opposite sign from that
of bosonic ones.

The total off-shell free energy of the scalar field
in the five-dimensional RN black hole is the
sum of those of bosonic and fermionic fields
\begin{eqnarray}
F_5 = &&
\sum_{i} F_5 (m_{B_i}) -
\sum_{i} F_5 ( m_{F_i})
\nonumber\\
= &&
\sum_{i} F_5^{{\rm div}} (m_{B_i}) -
\sum_{i} F_5^{{\rm div}} ( m_{F_i})
+ \sum_{i} F_5^{{\rm ren}} (m_{B_i}) -
\sum_{i} F_5^{{\rm ren}} ( m_{F_i}),
\end{eqnarray}
where $F_5 (m)$ is given by Eq. (\ref{fr5}).
It can be shown easily that the ultraviolet
divergences which consist of
$\frac{1}{\epsilon^{\frac{3}{2}}}$, $\frac{1}{\epsilon}$,
$\frac{1}{\epsilon^{\frac{1}{2}}}$, and $\ln (\epsilon)$,
may be removed, provided
that the number of bosonic and fermionic fields
are equal
\begin{equation}
N_B = N_F =3,
\label{num5}
\end{equation}
and the masses are required to satisfy
\begin{equation}
\sum_{i}^{3} m^2_{B_i} =
\sum_{i}^{3}  m^2_{F_i}.
\label{ma5}
\end{equation}
Note that these are the same conditions imposed
to regularize the one-loop effective action
in Sect. II.
Then the remaining matter field contribution
of the off-shell free energy
is simply given by
\begin{equation}
F_5 = F_5^{{\rm ren}}
=
 \frac{3 r_+}{16 \pi} \frac{\Gamma(2)
\zeta(2)}{(1-u) \beta^2} {{\cal B}_2},
\label{off fr5}
\end{equation}
where
\begin{equation}
{{\cal B}_2 } =
- \sum_{i=1}^{3} \ln (m^2_{B_i})
+
\sum_{i=1}^{3}  \ln (m^2_{F_i}).
\end{equation}
It should be noted that the matter
contribution (\ref{off fr5})
involves only ${\cal B}_2$,
which is the renormalization constant
(\ref{ren con1}) of six dimensions
related to the coupling constant.
The matter field contribution of the off-shell thermodynamic entropy,
$S = \beta^2 \frac{ \partial F}{\partial \beta}$,
is given by
\begin{equation}
S_5^{{\rm ren}}
= - \frac{3 r_+}{8 \pi} \frac{\Gamma(2)
\zeta(2)}{(1-u) \beta} {{\cal B}_2}.
\end{equation}
Substituting the Hawking temperature
\begin{equation}
\beta_H = \frac{4 \pi}{\Delta' (r_+)} =
\frac{2 \pi r_+}{1-u},
\end{equation}
one finds the matter field contribution of
the on-shell entropy
\begin{equation}
S_5^{{\rm ren}}
= - \frac{1}{32}
{{\cal B}_2}.
\end{equation}

We observe two important 
facts that are different from the four-dimensional case.
First of all the thermodynamic entropy does not
have a term proportional to the area
of black hole event horizon
\begin{equation}
A_5 = \frac{2 \pi^2 r_+^3}{\Gamma(2)},
\end{equation}
in five dimensions. It is remarkable that
the matter field contribution to the entropy in
five dimensions does
not involve the renormalization constant related
to the gravitational constant in strong contrast
with the four-dimensional case, in which
it was already observed that the renormalized
thermodynamic entropy of a scalar field in
the four-dimensional RN and the Schwarzschild black hole
contributes a quantum correction to the 
classical Bekenstein-Hawking entropy~\cite{demers,kim}. 
This means that in four dimensions
through the renormalization of the gravitational constant
the renormalization  constant ${\cal B}_1$ renormalizes
also the Bekenstein-Hawking entropy
\begin{equation}
\frac{A_4}{4G} + \frac{{\cal B}_1 A_4}{12 (4\pi)}
 = \frac{A_4}{4 G^{{\rm ren}}},
\end{equation}
so that the area-law of black hole entropy is still valid
even when one includes the matter field contribution of
thermodynamic entropy.
Secondly, we note that 
the matter field contribution of the thermodynamic
entropy has
the negative sign. The physical argument 
for these two facts is lacking at present.

We now turn to the six-dimensional RN black hole.
The free energy is again the sum of those of bosonic
and fermionic fields
\begin{equation}
F_6 =
\sum_{i} F_6 (m_{B_i}) -
\sum_{i} F_6 ( m_{F_i}).
\end{equation}
There are divergences proportional to
$\frac{1}{\epsilon^2}$, $\frac{1}{\epsilon}$,
and $\ln(\epsilon)$.
These ultraviolet divergences are removed by
the conditions
\begin{equation}
N_B = N_F = 4,
\end{equation}
and
\begin{equation}
\sum_{i=1}^{4} m_{B_i}^2 =
\sum_{i=1}^{4} m_{F_i}^2.
\end{equation}
Beside the ultraviolet divergent terms
there are also terms in Eq. (\ref{ren6})
which  become large as the masses
of regulator fields become large. These terms can be removed
by imposing an additional mass condition
\begin{equation}
\sum_{i=1}^{4} m^4_{B_i} = \sum_{i=1}^{4} m^4_{F_i}.
\label{ma6}
\end{equation}
Under these conditions satisfied, one obtains
the matter field contribution of the off-shell entropy
\begin{eqnarray}
S^{{\rm ren}}_6 = &&
 \frac{2r_+^5}{135 \pi}
\Biggl[
 \frac{15 \Gamma(2) \zeta(2)}{4(1-u) \beta}
 {{\cal B}_1}
 + \frac{(160-55u) \Gamma(4) \zeta(4) }{6(1-u)^3 \beta^3}
{{\cal B}_2}
\nonumber\\
+ &&
\frac{(352+1120u-635u^2)
\Gamma(6) \zeta(6)}{12(1-u)^5 \beta^5}
{{\cal B}_3} \Biggr]
\nonumber\\
+ &&
\frac{2r_+^3}{27 \pi}
\Biggl[
\frac{3 \Gamma(2) \zeta(2)}{(1-u) \beta}
{{\cal B}_2}
+ \frac{2(4 + u) \Gamma(4) \zeta(4)}{(1-u)^3 \beta^3}
{{\cal B}_3} \Biggr].
\label{off ren6}
\end{eqnarray}
Substituting the Hawking temperature
\begin{equation}
\beta_H = \frac{4 \pi r_+}{3 (1-u)},
\end{equation}
one obtains the on-sell entropy
\begin{eqnarray}
S^{{\rm ren}}_6 = &&
 \frac{\Gamma(2) \Gamma(\frac{5}{2})
 \zeta(2)}{12 \pi^{\frac{9}{2}}}
 {{\cal B}_1} \frac{A_6}{4}
\nonumber\\
+ && r_+^2
 \Biggl(\frac{(160-55u) \Gamma(4)
 \zeta(4) }{64 \pi^3}
 + \frac{\Gamma(2) \zeta(2)}{6 \pi^2} \Biggr)
{{\cal B}_2}
\nonumber\\
+ && \Biggl(
\frac{(352+1120u-635u^2)
\Gamma(6) \zeta(6)}{10240 \pi^6}
+  \frac{(4 + u) \Gamma(4) \zeta(4)}{16 \pi^3}
\Biggr)
{{\cal B}_3},
\label{on ren6}
\end{eqnarray}
where
\begin{equation}
A_6 = \frac{2 \pi^{\frac{5}{2}} r_+^4}{\Gamma(\frac{5}{2})}
\end{equation}
is the area of the six dimensional black hole event horizon.
In fact we obtain the renormalized off-shell (\ref{off ren6})
and on-shell (\ref{on ren6}) contribution of the 
thermodynamic entropy by the quantum matter field
which involves the same renormalization constants (\ref{ren con1})
used in the renormalization of the gravitational action.
The first term gives rise to a quantum correction to the
area-law of black hole.

Recollecting the renormalization of
the gravitational constant in Eq. (\ref{ren1})
in $D = 6$ $(n=3)$
\begin{equation}
\frac{1}{16 \pi G} + \frac{{\cal B}_1}{24 (4 \pi)^3}
= \frac{1}{16 \pi G^{\rm ren}},
\end{equation}
we see that the total entropy of the
scalar field, i.e the sum of the bare Bekenstein-Hawking
entropy and the quantum correction of entropy,
\begin{equation}
\frac{A_6}{4G} + \frac{{\cal B}_1 A_6}{24 (4 \pi)^2}
= \frac{A_6}{4 G^{\rm ren}}
\label{ren g}
\end{equation}
renormalizes the Bekenstein-Hawking entropy.
This implies that the Bekenstein-Hawking entropy
holds true at one-loop in $D = 6$ through
the renormalization of the gravitational constant.
The middle two terms proportional to $r_+^2$ are related with
the renormalization of one-loop quantum correction of
gravity. The last two terms renormalize
the cosmological constant.

\section{Discussion}

It should be remarked that the Pauli-Villars
regularization method with all the regulator fields
treated as obeying the Bose-Einstein or Fermi-Dirac
statistics
\begin{equation}
F_D = \mp \int_0^{\infty} dE \frac{g_D (E)}{e^{\beta E} \mp 1},
\end{equation}
depending on their spin-statistics~\cite{kim}
may not work for the RN black hole in higher than four
dimensions, because the ultraviolet divergent structure
has a hierarchy in which a higher dimensional
black hole has divergent terms peculiar to
that dimension and those also belonging to lower
dimensional black holes.
For instance in the five-dimensional RN black hole
the ultraviolet divergent part in (\ref{div5})
consists of terms proportional to $E^4$, $E^4$,
$E^2$, and $E$.
But the Pauli-Villars regularization method
with the correct spin-statistics
gives rise to the different statistical factors
from
\begin{eqnarray}
F_D &=&  \mp \int_0^{\infty} dE \frac{E^{k}}{e^{\beta E} \mp 1}
\nonumber\\
&=&
{\mit s}_k^{\mp}
\frac{\Gamma (k+1) \zeta (k+1)}{\beta^{k+1}},
\end{eqnarray}
where
\begin{eqnarray}
{\mit s}_k^{\mp} = \cases{1, \cr
 1 - \frac{1}{2^k}. \cr}
\end{eqnarray}
Thus to remove each divergent
term with a different power of $E$
one needs a different number of bosonic
and fermionic fields.
This means that all the ultraviolet divergent terms can not
be removed at the same time by the Pauli-Villars
regularization method with the correct spin-statistics.

\section{Conclusion}

In an attempt to understand how one might interpret
the matter field contribution to the black hole entropy,
we have studied the ultraviolet divergent structures of a massive
scalar field minimally coupled to a $D$-dimensional
Reissner-Nordstr\"{o}m black hole background
and computed its thermodynamic entropy
using the Pauli-Villars regularization method.
We have computed the entropy
with all the regulator fields treated as
bosons at finite temperature.
As explicit examples we have elaborated the five-
and six-dimensional cases in detail.

Interpreting the matter field contributions as the
one-loop contribution to the classical Bekenstein-Hawking
entropy, we compared the resultant
renormalization constants with those arising in the one-loop
renormalization of the gravitational action.
We have found that the matter field contribution does
not, in general, yield the correct renormalization
constants.
In particular, we have found that,
in an odd-dimensional spacetime,
the matter field contribution does not have a term
proportional to the surface area of the event horizon.
This result is consistent with the fact pointed out
by DLM~\cite{demers} that the correct interpretation
of the matter field contribution as renormalizing
the gravitational coupling constant,
is possible only when the scalar field coupling
is minimal, even in the four-dimensional spacetime.

\section*{acknowledgments}
This work was supported in parts by the Korea Science and
Engineering Foundation under Grant No. 951-0207-056-2
and 95-0701-04-01-3,
by the Basic Science Research Institute Program,
Ministry of Education
under Project No. BSRI-96-2418, BSRI-96-2425, BSRI-96-2427,
and by the Center for Theoretical
Physics, Seoul National University.

\appendix
\section{Integrals Used in Pauli-Villars Regularization}

We derive the integral formulas
useful in evaluating the effective action
of the scalar field using the Pauli-Villars
regularization method.

First in an even dimension,
we make use of the integral
\begin{equation}
\int_0^{\infty} dt e^{-zt} = \frac{1}{z}.
\label{A1}
\end{equation}
The integral (\ref{A1}) is well-defined
for a complex $z$ whose real part is positive,
but can be continued analytically even to a pure
imaginary $z$. We integrate the both sides of (\ref{A1})
with respect to $z$ from $\zeta_0$ to $z$ to obtain
\begin{equation}
\int_0^{\infty} dt \Bigl(\frac{e^{-zt}}{t}
- \frac{e^{- \zeta_0 t}}{t} \Bigr)  = - \ln (z) + \ln( \zeta_0),
\label{A2}
\end{equation}
and put $z =1$
\begin{equation}
\int_0^{\infty} dt \Bigl(\frac{e^{-t}}{t}
- \frac{e^{- \zeta_0 t}}{t} \Bigr) =  \ln( \zeta_0).
\label{A3}
\end{equation}
By subtracting (\ref{A3}) from (\ref{A2}) and rearranging,
we obtain
\begin{equation}
\int_0^{\infty} dt \frac{e^{-zt}}{t}
 = - \ln (z) + I_1,
\label{A4}
\end{equation}
where
\begin{equation}
I_1 = \int_0^{\infty} dt \frac{e^{-t}}{t}.
\label{A5}
\end{equation}
We repeat the integration of (\ref{A4}) with respect to
$z$ from $\zeta_1$ to $z$, we get
\begin{equation}
\int_0^{\infty} dt \Bigl(\frac{e^{-zt}}{t^2}
- \frac{e^{- \zeta_1 t}}{t^2} \Bigr) = z \ln (z)
- z - \zeta_1 \ln( \zeta_1) + \zeta_1 -
I_1(z - \zeta_1),
\label{A6}
\end{equation}
and put $z=0$ to the both sides of (\ref{A6}) to get
\begin{equation}
\int_0^{\infty} dt \Bigl( \frac{1}{t^2}
- \frac{e^{- \zeta_1 t}}{t^2} \Bigr) =
- \zeta_1 \ln( \zeta_1) + \zeta_1 + I_1 \zeta_1.
\label{A7}
\end{equation}
Subtract (\ref{A7}) from (\ref{A6}) to obtain
\begin{equation}
\int_0^{\infty} dt \frac{e^{-zt}}{t^2} = z \ln (z)
- z - z I_1 + I_2,
\label{A8}
\end{equation}
where
\begin{equation}
I_2 = \int_0^{\infty} dt \frac{1}{t^2}.
\label{A9}
\label{const}
\end{equation}
The $n$-times repetition of the integration
leads to the integral formula
\begin{equation}
\int_0^{\infty} dt \frac{e^{-zt}}{t^n} =
\frac{(-1)^n}{(n-1)!} z^{n-1} \ln (z)
+ \frac{(-1)^n}{(n-1)!} \bigl(I_1 + \sum_{k=1}^{n-1}
\frac{1}{k} \bigr) z^{n-1}
+ \sum_{l = 2}^{n} \frac{(-1)^{n-l}}{(n-l)!}
I_{l} z^{n-l},
\label{A10}
\end{equation}
where
\begin{equation}
I_p = \int_0^{\infty} dt \frac{1}{t^p},
\label{A11}
\end{equation}
for $p = 2, 3, \cdots$.

We can remove the infinite constants $I_p$
in (\ref{A10})
by adding and subtracting the integrals with different $z$.
For instance
\begin{equation}
\sum_{k=0}^{N} \int_0^{\infty} dt \Bigl(\frac{e^{-z_{2k} t}}{t} -
\frac{e^{-z_{2k+1} t}}{t} \Bigr) = -
\sum_{k=0}^{N} \Bigl(\ln(z_{2k})  - \ln(z_{2k+1}) \Bigr).
\label{A12}
\end{equation}
and
\begin{equation}
\sum_{k=0}^{N} \int_0^{\infty} dt \Bigl(\frac{e^{-z_{2k} t}}{t^2} -
\frac{e^{-z_{2k+1} t}}{t^2} \Bigr) =
\sum_{k=0}^{N} \Bigl(z_{2k}\ln(z_{2k})
- z_{2k+1} \ln(z_{2k+1}) \Bigr)
- I_1 \sum_{k=0}^{N} \Bigl(z_{2k} - z_{2k+1} \Bigr).
\label{A13}
\end{equation}
We impose a condition
\begin{equation}
\sum_{k=0}^{N} z_{2k} = \sum_{k=0}^{N} z_{2k+1}
\label{A14}
\end{equation}
to remove $I_1$.
In this way we obtain
\begin{equation}
\sum_{k=0}^{N} \int_0^{\infty} dt \Bigl(\frac{e^{-z_{2k} t}}{t^n} -
\frac{e^{-z_{2k+1} t}}{t^n} \Bigr) =
\frac{(-1)^n}{(n-1)!}
\sum_{k=0}^{N} \Bigl(z_{2k}^{n-1} \ln(z_{2k})
- z_{2k+1}^{n-1} \ln(z_{2k+1}) \Bigr),
\label{A15}
\end{equation}
provided that
\begin{equation}
\sum_{k=0}^{N} z_{2k}^m = \sum_{k=0}^{N} z_{2k+1}^m
\label{A16}
\end{equation}
for $m = 1, 2, \cdots, n-1$.
This is the main idea to get rid of the infinite quantities
in the Pauli-Villars regularization method.
We now may continue the integrals analytically
even to a complex $t$ bearing in mind the subtraction
procedure.

We can also obtain (\ref{A15}) without introducing
the infinite constants $I_p$. We add and subtract
(\ref{A2}) with different $z$ to get directly (\ref{A12}).
We integrate again (\ref{A2}) with respect to $z$
from $\zeta_1$ to $z$ to get
\begin{equation}
\int_0^{\infty} dt \Bigl(\frac{e^{-zt}}{t^2}
- \frac{e^{- \zeta_1 t}}{t^2} \Bigr)
+ (z - \zeta_1) \int_0^{\infty} dt
\frac{e^{- \zeta_1 t}}{t}
=  z\ln (z) - z \bigl(1 + \ln(\zeta_0) \bigr)
- \zeta_1 \bigl(\ln(\zeta_1) - \ln( \zeta_0)
-1 \bigr).
\label{A17}
\end{equation}
By adding (\ref{A17}) and subtracting with different $z$ we obtain
\begin{equation}
\sum_{k=0}^{N} \int_0^{\infty} dt \Bigl(\frac{e^{-z_{2k} t}}{t^2} -
\frac{e^{-z_{2k+1} t}}{t^2} \Bigr) =
\sum_{k=0}^{N} \Bigl(z_{2k}\ln(z_{2k})
- z_{2k+1} \ln(z_{2k+1}) \Bigr)
- \sum_{k=0}^{N} \Bigl(z_{2k} - z_{2k+1} \Bigr)
\bigl(1 + \ln(\zeta_0) \bigr).
\label{A18}
\end{equation}
We impose the condition (\ref{A14}) to get
\begin{equation}
\sum_{k=0}^{N} \int_0^{\infty} dt \Bigl(\frac{e^{-z_{2k} t}}{t^2} -
\frac{e^{-z_{2k+1} t}}{t^2} \Bigr) =
\sum_{k=0}^{N} \Bigl(z_{2k}\ln(z_{2k})
- z_{2k+1} \ln(z_{2k+1}) \Bigr).
\label{A19}
\end{equation}
We can repeat the procedure $n$-times to get
(\ref{A15}) provided that we impose the conditions
(\ref{A16}). In either way we get the identical
regularized integrals.

Similarly in an odd dimension,
integrating both sides of the integral
with respect to $z$
\begin{equation}
\int_0^{\infty} dt \frac{e^{-zt}}{t^{1/2}} =
\frac{\pi^{1/2}}{z^{1/2}},
\label{A20}
\end{equation}
we obtain
\begin{equation}
\int_0^{\infty} dt \frac{e^{-zt}}{t^{3/2}} =
- 2 \pi^{1/2} z^{1/2} + I_{3/2}.
\label{A21}
\end{equation}
We repeat the integration with respect to $z$
to get
\begin{equation}
\int_0^{\infty} dt \frac{e^{-zt}}{t^{5/2}} =
\frac{4 \pi^{1/2}}{3} z^{3/2} - I_{3/2}z
+ I_{5/2}.
\label{A22}
\end{equation}
Repeating $n$-times we get
\begin{equation}
\int_0^{\infty} dt \frac{e^{-zt}}{t^{(2n+1)/2}} =
\frac{(-1)^n 2^{2n} \pi^{1/2} n!}{(2n)!} z^{(2n-1)/2}
+ \sum_{l = 1}^{n} \frac{(-1)^{n-l}}{(n-l)!}
I_{(2l+1)/2} z^{n-l}.
\label{A23}
\end{equation}

\section{Integral Formula}

In calculating the free energy
we used frequently the following integral formula~\cite{grad}
for integral $q$ and $p$:
\begin{equation}
\int dx x^q Z^{\frac{2p - 1}{2}}
\end{equation}
where $Z (x) = E^2 - (1 - u)m^2 x$.
For positive integers $q$ and $p$
it becomes
\begin{eqnarray}
A^{q}_{\frac{2p-1}{2}} = &&\int_\epsilon dx
\frac{Z^{\frac{2p - 1}{2}} (x)}{x^q}
=  Z^{\frac{2p+1}{2}} (\epsilon)
\Biggl(
\frac{1}{(q-1) E^2 \epsilon^{q-1}}
 + \sum_{l = 1}^{q-2} (-1)^l
 \nonumber\\
&& \times
\frac{(2p-2q + 3)(2p-2q+5) \cdots (2p-2q + 2l+1)}{
2^l (q-1) (q-2) \cdots (q-l-1) \epsilon^{q-l-1}}
\frac{1}{E^2} \Bigl( \frac{(1-u)m^2}{E^2} \Bigr)^{l}
\Biggr)
\nonumber\\
&& +  (-1)^{q}
\frac{(2p-2q + 3)(2p-2q+5) \cdots (2p-3)(2p-1)}{
2^{q-1} (q-1)!}
\nonumber\\
&& \times
\Bigl(\frac{(1-u)m^2}{E^2} \Bigr)^{q-1}
\int_\epsilon dx \frac{Z^{\frac{2p - 1}{2}} (x)}{x},
\nonumber\\
A^1_{\frac{2p-1}{2}} = && \int_\epsilon dx
\frac{Z^{\frac{2p - 1}{2}} (x)}{x}
=  -  2 \sum_{l=1}^{p}
\frac{E^{2(p-l)} Z^{\frac{2l-1}{2}}(\epsilon) }{2l-1}
- E^{2p-1}
\ln \Bigl(\frac{E - Z^{\frac{1}{2}} (\epsilon)}{
E + Z^{\frac{1}{2}} (\epsilon)} \Bigr),
\nonumber\\
A^1_{\frac{1}{2}} = && \int_\epsilon dx \frac{1}{x Z^{\frac{1}{2}} (x)}
=  -
\frac{1}{E}
\ln \Bigl(\frac{E - Z^{\frac{1}{2}} (\epsilon)}{
E + Z^{\frac{1}{2}} (\epsilon)} \Bigr).
\end{eqnarray}
and also for positive integers $q$ and $p$
\begin{equation}
\int_\epsilon dx x^q Z^{\frac{2p - 1}{2}} (x)
= (-1)^q 2 \Bigl(\sum_{l=0}^{q}
\frac{(-1)^l {q \choose l} E^{2l}
Z^{q - l}(\epsilon) }{
2q - 2l + 2p + 1} \Bigr)
\frac{Z^{\frac{2p+1}{2}} (\epsilon) }{
\bigl( (1 - u)m^2 \bigr)^{q +1}}.
\end{equation}
For integral $q$ and $p$ we also have
\begin{eqnarray}
\int_\epsilon dx \frac{x^q}{Z^{\frac{p}{2}} (x)}
= &&
\frac{\epsilon^q}{ \bigl(1- \frac{p}{2}\bigr)
(1 -u) m^2  Z^{\frac{p}{2}-1} (\epsilon) }
+ \frac{q}{ \bigl(1- \frac{p}{2} \bigr)(1-u)m^2 }
\int_\epsilon dx \frac{x^{q-1}}{Z^{\frac{p}{2} -1} (x)},
\nonumber\\
\int_\epsilon dx \frac{1}{Z^{\frac{p}{2}} (x)}
= &&
\frac{1}{\bigl(1- \frac{p}{2}\bigr)
(1 -u) m^2  Z^{\frac{p}{2}-1} (\epsilon)}
\end{eqnarray}
For $\epsilon << 1$, one has $Z(\epsilon) = E^2
+ O(\epsilon)$. Keeping only divergent and nonvanishing terms
as $\epsilon$ goes to zero,
we get the following integrals frequently used in five and six
dimensions
\begin{eqnarray}
A^1_{\frac{1}{2}} = &&
- 2E - E \ln \Bigl(\frac{(1-u)m^2}{4 E^2} \epsilon \Bigr),
\nonumber\\
A^2_{\frac{1}{2}} = &&
 \frac{E}{\epsilon}
 + \frac{(1-u)m^2}{E}
 + \frac{(1-u)m^2}{2E} \ln \Bigl(\frac{(1-u)m^2}{4 E^2}
 \epsilon \Bigr),
 \nonumber\\
A^3_{\frac{1}{2}} = &&
 \frac{E}{2\epsilon^2}
 + \frac{(1-u)m^2}{4E \epsilon}
 + \frac{(1-u)^2 m^4}{4 E^3}
 + \frac{(1-u)^2m^4}{8E^3} \ln \Bigl(\frac{(1-u)m^2}{4 E^2}
 \epsilon \Bigr),
 \label{}
\end{eqnarray}
and
\begin{eqnarray}
A^1_{\frac{3}{2}} = &&
- \frac{8E^3}{3} - E^3 \ln \Bigl(\frac{(1-u)m^2}{4 E^2} \epsilon \Bigr),
\nonumber\\
A^2_{\frac{3}{2}} = &&
 \frac{E^3}{\epsilon}
 + 4 (1-u)m^2 E
 + \frac{3(1-u)m^2 E}{2} \ln \Bigl(\frac{(1-u)m^2}{4 E^2}
 \epsilon \Bigr),
 \nonumber\\
A^3_{\frac{3}{2}} = &&
 \frac{E^3}{2\epsilon^2}
 - \frac{(1-u)m^2 E}{4 \epsilon}
 - \frac{(1-u)^2 m^4}{E}
 - \frac{3(1-u)^2m^4}{8E} \ln \Bigl(\frac{(1-u)m^2}{4 E^2}
 \epsilon \Bigr),
 \label{}
\end{eqnarray}
and
\begin{eqnarray}
A^1_{\frac{5}{2}} = &&
- \frac{46E^5}{15}
- E^5 \ln \Bigl(\frac{(1-u)m^2}{4 E^2} \epsilon \Bigr),
\nonumber\\
A^2_{\frac{5}{2}} = &&
 \frac{E^5}{\epsilon}
 + \frac{23 (1-u)m^2 E^3}{3}
 + \frac{5(1-u)m^2 E^3}{2}
 \ln \Bigl(\frac{(1-u)m^2}{4 E^2}
 \epsilon \Bigr),
 \nonumber\\
A^3_{\frac{5}{2}} = &&
 \frac{E^5}{2\epsilon^2}
 - \frac{3(1-u)m^2 E^3}{4 \epsilon}
 - \frac{23(1-u)^2 m^4E}{4}
 - \frac{15(1-u)^2m^4 E}{8}
 \ln \Bigl(\frac{(1-u)m^2}{4 E^2}
 \epsilon \Bigr),
 \label{}
\end{eqnarray}
and
\begin{eqnarray}
A^1_{\frac{7}{2}} = &&
- \frac{352E^7}{105}
 - E^7 \ln \Bigl(\frac{(1-u)m^2}{4 E^2} \epsilon \Bigr),
\nonumber\\
A^2_{\frac{7}{2}} = &&
 \frac{E^7}{\epsilon}
 + \frac{176 (1-u)m^2 E^5}{15}
 + \frac{7(1-u)m^2 E^5}{2}
 \ln \Bigl(\frac{(1-u)m^2}{4 E^2}
 \epsilon \Bigr),
 \nonumber\\
A^3_{\frac{7}{2}} = &&
 \frac{E^7}{2\epsilon^2}
 - \frac{5(1-u)m^2 E^5}{4 \epsilon}
 - \frac{44(1-u)^2 m^4 E^3}{3}
 - \frac{35(1-u)^2m^4 E^3}{8}
 \ln \Bigl(\frac{(1-u)m^2}{4 E^2}
 \epsilon \Bigr),
 \label{}
\end{eqnarray}
and
\begin{eqnarray}
A^1_{\frac{9}{2}} = &&
- \frac{1126E^9}{315}
 - E^9 \ln \Bigl(\frac{(1-u)m^2}{4 E^2} \epsilon \Bigr),
\nonumber\\
A^2_{\frac{9}{2}} = &&
 \frac{E^9}{\epsilon}
 + \frac{563 (1-u)m^2 E^7}{35}
 + \frac{9(1-u)m^2 E^7}{2}
 \ln \Bigl(\frac{(1-u)m^2}{4 E^2}
 \epsilon \Bigr),
 \nonumber\\
A^3_{\frac{9}{2}} = &&
 \frac{E^9}{2\epsilon^2}
 - \frac{7(1-u)m^2 E^7}{4 \epsilon}
 - \frac{563(1-u)^2 m^4 E^5}{20}
 - \frac{63(1-u)^2m^4 E^5}{8}
 \ln \Bigl(\frac{(1-u)m^2}{4 E^2}
 \epsilon \Bigr).
 \label{}
\end{eqnarray}


\begin{references}

\bibitem{bekenstein} J. D. Bekenstein, Phys. Rev.
D {\bf 9}, 3292 (1974).
\bibitem{hawking} S. W. Hawking, Commun. Math. Phys.
{\bf 43}, 199 (1975).
\bibitem{thooft} 't Hooft, Nucl Phys. {\bf B256}, 727 (1985).
\bibitem{susskind} L. Susskind and J. Uglum, Phys.
Rev. D {\bf 50}, 2700 (1994).
\bibitem{demers} J-G. Demers, R. Lafrance,
and R. C. Myers, Phys. Rev. D {\bf 52}, 2245 (1995).
\bibitem{mann} R. B. Mann, L. Tarasov, and
A. Zelnikov, Class. Quantum Grav. {\bf 9}, 1487 (1992).
\bibitem{bytsenko} A. A. Bytsenko, G. Cognola,
and S. Zerbini, Nucl Phys. B {\bf 458}, 267 (1996).
\bibitem{birrel} N. D. Birrel and P. C. W. Davies,
{\it Quantum Fields in Curved Spacetime} (Cambridge
University Press, Cambridge, 1982).
\bibitem{dewitt} B. S. DeWitt, Phys. Rep. {\bf 19}, 295 (1975).
\bibitem{myers} R. Myers and M. Perry, Ann. Phys. {\bf 172},
304 (1986).
\bibitem{grad} {\it Table of Integral, Series, and Products}
edited by I. S. Gradsheteyn and I. M. Ryzhik
(Academic Press, New York, 1980).
\bibitem{kim} S. P. Kim, S. K. Kim,
K.-S. Soh, and J. H. Yee, "Remarks
on Renormalization of Black Hole Entropy",
Seoul National University Preprint,
SNUTP-96032, gr-qc/9607019 (1996).
\end{references}
\end{document}